\documentclass[12pt]{iopart}
\usepackage{times}

\usepackage{iopams}
\usepackage{graphicx}
\usepackage{bm}
\usepackage{bbm}
\usepackage{color}
\usepackage{verbatim}
\usepackage{tikz}
\usepackage{hyperref}
\usepackage{float}
\usepackage{harvard}

\begin{document}

\title{Bandwidth-resonant Floquet states in honeycomb optical lattices}

\author{A. Quelle$^1$, M. O. Goerbig$^2$, and C. Morais Smith$^1$}

\address{$^1$ Institute for Theoretical Physics, Center for Extreme Matter and Emergent Phenomena, Utrecht University, Leuvenlaan 4, 3584 CE Utrecht, The Netherlands}
\address{$^2$ Laboratoire de Physique des Solides, CNRS UMR 8502, Universit\'e Paris-Sud, B\^at. 510,
F-91405 Orsay cedex}

\begin{abstract}

We investigate, within Floquet theory, topological phases in the out-of-equilibrium system that consists of fermions in a circularly shaken honeycomb optical lattice. We concentrate on the intermediate regime, in which the shaking frequency is of the same order of magnitude as the band width, such that adjacent Floquet bands start to overlap, creating a hierarchy of band inversions. It is shown that two-phonon resonances provide a topological phase that can be described within the Bernevig-Hughes-Zhang model of HgTe quantum wells. This allows for an understanding of out-of-equilibrium topological phases in terms of simple band inversions, similar to equilibrium systems.

\end{abstract}

\maketitle

\section{Introduction}

The recent realisation of topoplogical states in condensed-matter systems \cite{Konig2007,Hasan2010,Zhang2010} has sparked a flurry of activities in the field of cold atoms, aiming at reproducing, engineering, and manipulating these fascinating quantum states in traps \cite{Roncaglia2011,Corman2014,Burrello2015} and in optical lattices \cite{Hemmerich2007,Aidelsburger2013,Miyake2013,Jotzu2014,Aidelsburger2015,Chomaz2015}.

Topological fermionic systems exhibit protected metallic states at the boundary, while the bulk of the material remains insulating. In cold atoms, quantised transverse density currents play the role of the quantised charge currents, since we deal with neutral atoms in selected hyperfine states, instead of spin-full electrons. 
Similarly, the pair of atomic hyperfine states yields a spin-1/2 structure that allows for the analogue of quantised spin currents obtained in condensed-matter systems. The conductivity at the boundary is given by a topological invariant, which is quantised and stable against perturbations. 

There are a variety of topological states known by now, and they are well described by the ten-fold classification \cite{Ryu2010}, which defines a topological invariant according to the symmetries and dimensionality of the system. Although well established, the ten-fold classification neglects several effects. First, it does not take into account interactions, which leads to the fractional quantum Hall effect \cite{Tsui1982,Laughlin1983}, to the quantum anomalous Hall effect \cite{Nandkishore2010,Jung2011}, or as recently shown, to a quantum valley Hall effect \cite{Marino2015}. Second, it does not consider the crystal symmetry of the lattice, which may give rise to crystalline topological insulators, and even more general behaviour \cite{Slager2013}. Last, but not least, it does not take into account out-of-equilibrium systems. 

The case of topological insulators (TI's) under the influence of a time-periodic perturbation, the so-called Floquet TI's (FTI's) \cite{Kitagawa2011,Inoue2010,Gu2011,Lindner2011,Ezawa2013,Fregoso2013,Rudner2013,Wang2013,Fregoso2014,Gomez-Leon2014PRB,Perez-Piskunow2014,Quelle2014QSH,Kundu2014,Usaj2014,Carpentier2015}, has, until now, been considered in three unequal regimes. Firstly the so-called quasi-equilibrium regime, where $J\ll\hbar\omega\ll\Delta$; here $J$ is the hopping parameter, which is roughly the bandwidth of the relevant set of bands, $\omega$ the driving frequency, and $\Delta$ the gap between the next set of bands. In this case, the system constituents cannot follow the perturbation, and the system remains at quasi-equilibrium with simply renormalised lattice parameters. It is the regime that has been most studied \cite{Eckardt2005,Arimondo2009,Koghee2012,Sengstock2013,Goldman2014}. Secondly, the regime where $J\ll\hbar\omega\sim \Delta.$ This regime is starting to attract interest in optical lattices \cite{Cheng2013,Zhai2014,Zhai2014PRA}, but it has been unexplored in the context of condensed matter. Thirdly, the regime where $J\sim \hbar\omega\ll\Delta$, and where the equilibrium topological classification breaks down. It is this regime where most of the work on FTI's in condensed matter has taken place, and these kind of systems have even been simulated in twisted photonic waveguides \cite{Rechtsman2013}, where the third spatial dimension takes the role of time. There have been attempts to define Chern-type topological invariants valid for every frequency range \cite{Lindner2011,Rudner2013,Carpentier2015}, and it is known that these invariants reduce to the equilibrium ones in the first regime. The transition between the first and third regime has been investigated theoretically in Ref.~\cite{Kundu2014} for graphene irradiated by circularly polarised light. However, for the case of ultracold atoms, this regime has so far been overlooked.

Here, we show that for circularly shaken honeycomb optical lattices the transition between the first and third regime can be understood in terms of band inversion. These band inversions occur because, with decreasing frequency, the different Floquet bands start to overlap. Because the circular shaking induces phonon resonances, avoided crossings that generally host topological edge states occur. The polarisation of the shaking breaks time-reversal symmetry, such that the resulting FTI is in the quantum anomalous Hall class. It has recently been shown that one-phonon resonances create an additional topological gap in the spectrum at non-zero energy, whereas two-phonon resonances destroy the topological nature of the zero-energy gap by creating counter-propagating edge states \cite{Quelle2014QSH}. Here, we derive an effective continuum model from the out-of-equilibrium lattice model, and show that, in the vicinity of the band inversion occurring for the two-phonon resonances, it turns out to be described by the Bernevig-Hughes-Zhang (BHZ) model. Additionally, we demonstrate the validity of this model by comparing our results with a numerical solution of the full problem. In this way, our results describe the transition between the quasi-equilibrium and the resonant regimes of the shaken system, by modelling the appearance of phonon resonances in the latter regime in terms of band inversions. This provides a link between non-equilibrium and equilibrium topological states of matter. Finally, we discuss a possible experimental observation in shaken honeycomb optical lattices loaded with ultracold fermions, as well as in the recently realised honeycomb superlattices of CdSe nanocrystals \cite{Kalesaki2014,Bone2014}.

\section{The Hamiltonian in co-moving coordinates}

Consider an optical lattice that is shaken in time. The deviation of the lattice from its equilibrium position is denoted by $\bm r(t)$; by assumption $\bm r(t+T)=\bm r(t)$ for some period $T$. To find the Hamiltonian in co-moving coordinates, we consider the Poincar\'e-Cartan form
\begin{eqnarray}
dS=\bm p\cdot d\bm q-Hdt
\end{eqnarray}
along the trajectory of a system in phase space.
We change coordinates to co-moving coordinates $\bm{\tilde q}=\bm q+\bm r(t)$, $\bm{\tilde p}=\bm p$, such that $d\bm{\tilde p}=d\bm p$ and $d\bm{\tilde q}=d\bm q+\dot{\bm r}(t)dt$, where $\dot{\bm r}(t)$ is the time derivative of $\bm r(t)$. The Poincar\'e-Cartan form can thus be rewritten as
\begin{eqnarray}
dS&=\bm p\cdot d\bm q-Hdt=\bm{\tilde p}\cdot[d\bm{\tilde q}-\dot{\bm r}(t)dt]-Hdt\\
&=\bm{\tilde p}\cdot d\bm{\tilde q}-\left(H+\bm{\tilde p}\cdot\dot{\bm r}(t) \right)dt.
\end{eqnarray}
We immediately read off the Hamiltonian in co-moving coordinates: $\tilde H=H+\bm{\tilde p}\cdot \dot{\bm r}(t)$. The extra term encodes the pseudoforces seen because the co-moving frame is not inertial. For a shaken optical lattice, the Hamiltonian reads
\begin{eqnarray}\label{Hamiltonian}
H=\frac{p^2}{2m}+V(\bm q+\bm r(t)),
\end{eqnarray}
where the potential determines the lattice, which we take to be honeycomb.

In the co-moving frame $\bm{\tilde q}$ and $\bm{\tilde p}$, Eq.~\eref{Hamiltonian} becomes
\begin{eqnarray}\label{CHamiltonian}
\tilde H&=\frac{\tilde p^2}{2m}+V(\bm{\tilde q})+ \bm{\tilde p}\cdot \dot{\bm r}(t)\\
&=\frac{\|\bm{\tilde p}+m \dot{\bm r}(t)\|^2}{2m}+V(\bm{\tilde q})-\frac{1}{2}m \|\dot{\bm r}(t)\|^2.
\end{eqnarray}
For circular shaking, $\bm r(t)=r_0 (\cos(\omega t),\sin(\omega t)),$ which means $\|\dot{\bm r}(t)\|^2$ is constant in time and can be ignored by shifting the energy. The circular shaking thus induces a rotating vector potential $e \bm A$, which has constant magnitude $eA=m r_0\omega$. Compare this to the Hamiltonian for a system irradiated by circularly polarised light, where $eA=eE/\omega$ for electric field $E$. 

\section{Floquet theory}
The Hamiltonian in Eq.~\eref{CHamiltonian} is periodic in time, so according to Floquet theory \cite{Sambe1972,Hemmerich2010}, the time-dependent Schr\"odinger equation has quasi-periodic solutions $\psi(t)=\exp\left(-i\epsilon t/\hbar\right)\phi(t)$, where $\phi$ is a periodic function in time and thus a solution of $H_F\phi(t)=\epsilon \phi(t).$ Here, the Floquet Hamiltonian is defined as $H_F:=H-i\hbar\partial_t.$ If $H$ acts on the Hilbert space ${\mathcal{H},}$ and ${\mathcal{H}_T}$ is the Hilbert space of $T$-periodic functions, then $H_F$ acts on ${\mathcal{H}\otimes\mathcal{H}_T}$. The space ${\mathcal{H}_T}$ is spanned by the functions $|n\rangle:=\exp(i n\omega t)$, and has inner product $1\int dt/T$. With respect to the states $|n\rangle,$ we can write $H_F$ as a block matrix with elements
\begin{eqnarray}\label{FloqHamElts}
 \nonumber
\langle n|H_F|m \rangle &=& \frac{1}{T}\int dt \exp\left[i\omega(m-n)t\right](H+m\hbar\omega)\\
&=& :H_{m-n}+m\hbar\omega\delta_{m,n}. 
\end{eqnarray}
Here, the $H_{m-n}$ are the Fourier modes of the original Hamiltonian $H$.


\section{Model}
\begin{figure}[b]
\centering
\includegraphics[width=0.5 \linewidth]{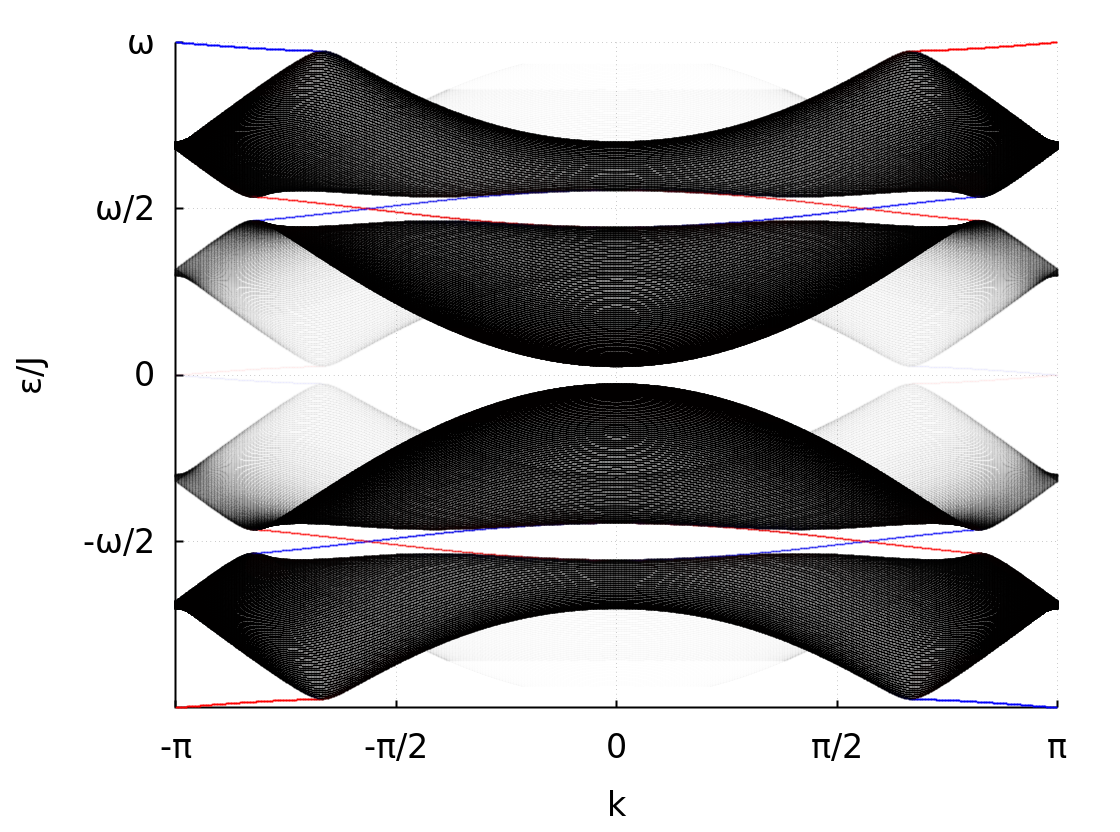}
\caption{\label{graph30} (Colour online) The spectrum of the Floquet Hamiltonian $H_F$ is shown. Plots were made for a ribbon geometry with zigzag edges, and $k$ denotes the Bloch momentum along the length of the ribbon. Two periods of the spectrum of $H_F$ are shown for $\hbar \omega=3 J$ and $m r_0 \omega^2 a=J.$ The relevant feature is the impending gap closure at $\epsilon=0$ and $k=0$, when the Floquet bands $n=1$ and $n=-1$ overlap. To highlight this, we have made all bands, except for $n=1$ and $n=-1$, translucent. Red and blue represent different edges; the entire spectrum is spin degenerate.}
\end{figure}

\begin{figure*}[t]
\includegraphics[width=\linewidth]{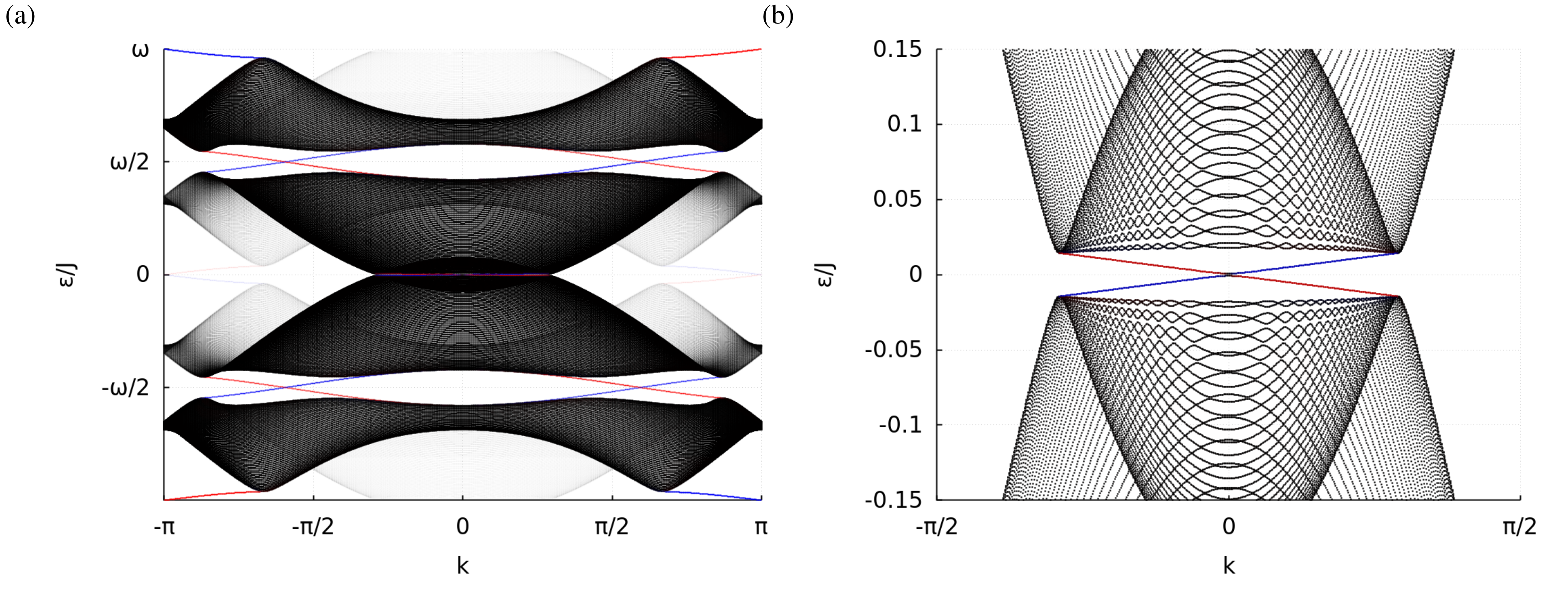}
\caption{\label{graph27} (Colour online) (a) Same as in Fig.~\ref{graph30}, but for $\hbar \omega=2.7 J$. The relevant feature is the band inversion at $\epsilon=0$ and $k=0$, where the Floquet bands $n=1$ and $n=-1$ overlap, creating a gap with topologically protected edge states. To highlight this, we have made all bands, except for $n=1$ and $n=-1$, translucent. (b) A zoom in on the band crossing from (a) is provided to make the details of the gap visible.}
\end{figure*}

We apply the Floquet formalism to fermions in a circularly shaken honeycomb lattice, with shaking radius $r_0$, and frequency $\omega$. We work in the co-moving reference frame, where the Hamiltonian has the form in Eq.~\eref{CHamiltonian}. To facilitate our analysis we use the tight-binding approximation, as done for graphene \cite{CastroNeto2009}. In second quantisation, 
the result is the Bloch Hamiltonian, except that one has to account for the vector potential according to the Peierls substitution, ${\bm k\mapsto \tilde{\bm k}:=\bm k+e \bm A/\hbar}:$
\begin{eqnarray}\label{Ham}
H(\bm k,t)=
J\sum_l\left( \begin{array}{cc}
0 & \exp(i \tilde{\bm k}\cdot \bm \delta_l)  \\
 \exp(-i \tilde{\bm k}\cdot \bm \delta_l) & 0
\end{array} \right).
\end{eqnarray}
Here, $J>0$ is the NN hopping amplitude, and our convention for the NN hopping vectors $\delta_l$ is $\bm \delta_0 = a(0,1),$ and ${\bm \delta_{\pm 1}=- a\left(\pm\sqrt{3},1\right)/2},$ where $a$ is the NN distance. Consequently, one obtains, via Eq.~\eref{FloqHamElts}, the matrices $H_n$ 
\begin{equation}\label{Hn}
\fl H_n=J \left( \begin{array}{cc}
0 & \sum_l\kappa_{-n}\exp[i\left(\bm k\cdot \bm \delta_l+\alpha_{l,n}\right)]  \\
\sum_l\kappa_n\exp[i\left(\bm -k\cdot \bm \delta_l+\alpha_{l,n}\right)] & 0
\end{array} \right).
\end{equation}
In Eq.~\eref{Hn}, $\kappa_n:=J_n\left(a m r_0\omega/\hbar\right),$ where $J_n$ is the Bessel function of the $n^{th}$ kind, and $\alpha_{l,n}:=n {\rm Arg}[\bm \delta_l]+n\pi/2$, where $\rm{Arg}$ gives the angle of a vector with the $x$-axis.  Using Eq.~\eref{Hn}, we obtain $H_F$ as an infinite block matrix
\begin{equation}\label{blockFloqHam}
H_F=\left( \begin{array}{ccccc}
\ddots&\vdots&\vdots&\vdots&\vdots\\
\cdots&H_0+\hbar\omega& H_1&H_2&\cdots\\
\cdots&H_{-1}& H_0&H_1&\cdots\\
\cdots&H_{-2}& H_{-1}&H_0-\hbar\omega&\cdots\\
\vdots&\vdots&\vdots&\vdots&\ddots\\
\end{array} \right).
\end{equation} 
If the spectrum around an energy $n\hbar\omega$ is desired, it can be computed by truncating the right-hand side of Eq.~\eref{blockFloqHam} around $H_0+n\hbar\omega.$ 
By tuning the shaking frequency $\omega$, one may reach certain regimes where $H_F$ exhibits topological edge states. In the large-frequency limit, $\hbar \omega \gg J$, the Floquet bands are well separated, but one finds a hierarchy of band crossings upon decrease of $\omega$, when $\hbar\omega\approx J$. These band-crossings create additional, possibly topological gaps in the spectrum. In the following, we focus on the band crossing in the vicinity of $k=0$ and $\epsilon=0$, appearing in the interesting regime: $\hbar \omega\approx 2.9J$. In Fig.~\ref{graph30}, the spectrum of $H_F$ is plotted for $\hbar\omega=3J$. The bottom of the valence band from $H_0+\hbar\omega$ (on top) and the top of the conduction band from $H_0-\hbar\omega$ (below) are visible with a gap between them. As $\omega$ is lowered, the valence band (on top) descends and the conduction band (on bottom) ascends; a band inversion takes place at $\hbar \omega\approx 2.9J$, creating an avoided crossing [Fig.~\ref{graph27}(a)]. At the avoided crossing, a single pair of edge states crosses the gap, as is highlighted in Fig.~\ref{graph27}(b), where we show a zoom in on a narrow energy window. It should be noted that the edge states at $k=0$, which correspond to the two-phonon resonance, are counter-propagating with respect to the zero-phonon resonance edge states at the Dirac points. (These also occur in the gap at $\varepsilon=0$, but have been made translucent to avoid confusion. They are depicted at $\varepsilon=\pm \hbar\omega$, which are equivalent to $\varepsilon=0$ due to the periodicity of the Floquet spectrum; in these gaps, instead, the two-phonon resonance states have been made translucent.) The appearance of new edge states at $k=0$ removes the topological protection of the edge states in the $\varepsilon=0$ gap \cite{Quelle2014QSH}. 

This lack of topological protection can be better verified in a lattice with armchair termination, where the zero-phonon and two-phonon resonances both occur at $k=0$. Indeed, they gap out because of hybridisation. Furthermore, if there is a domain wall in the system, where the orientation of the irradiation changes, the states localised at the domain wall also gap out due to hybridisation effects \cite{Quelle2014FM}. This shows that the appearance of the two-phonon resonance indeed makes the $\varepsilon=0$ gap trivial. The situation can be reversed by applying a staggered sublattice potential: this will destroy the zero-phonon resonance at the Dirac points, but leave the two-phonon resonance untouched. In this case, the appearance of the two-phonon resonance changes the gap at $\varepsilon=0$ from topologically trivial to non-trivial. 


\section{Low-energy effective theory}
To write down an effective theory that allows us to characterise the band crossing, the gap and the dispersion of the edge states, we extract the relevant energy bands from $H_F$. This is done by diagonalizing $H_0$, and for simplicity we keep terms up to second order in $\bm k$. We define the unitary transformation
\begin{equation}\label{basischange}
U:=\frac{1}{2}\sqrt{2}\left( \begin{array}{cc}
1&1\\
1&-1
\end{array} \right)
\end{equation}
and consider the transformed Hamiltonian
\begin{equation}\label{UFloqHam}
\tilde H_F=\left( \begin{array}{ccccc}
\ddots&\vdots&\vdots&\vdots&\vdots\\
\cdots&\tilde H_0+\hbar\omega& \tilde H_1&\tilde H_2&\cdots\\
\cdots&\tilde H_{-1}& \tilde H_0&\tilde H_1&\cdots\\
\cdots&\tilde H_{-2}& \tilde H_{-1}&\tilde H_0-\hbar\omega&\cdots\\
\vdots&\vdots&\vdots&\vdots&\ddots\\
\end{array} \right),
\end{equation}
where ${\tilde H_n=U H_n U}$ by construction. From Eqs.~\eref{Hn}~and~\eref{basischange}, one finds the identity 
\begin{equation*}
\tilde H_0=J \kappa_0\left(3-\frac{3}{4}k^2\right)\sigma_z +...
\end{equation*}
It follows that for $\hbar\omega\approx 3J$ and $\bm k\approx0,$ the $(2,2)$ matrix element of the matrix $\tilde H_0+\hbar\omega$ and the $(1,1)$ element of $\tilde H_0-\hbar\omega$ are much smaller than all the other energy scales in the problem. They are the zeroth-order energies of the two bands near the band crossing shown in Figs.~\ref{graph30} and \ref{graph27}. This leads us to define a second unitary transformation $V$ that is characterised by the matrix elements $\langle n|V|m\rangle = \sigma_+\delta_{m,n-1} + \sigma_-\delta_{m,n+1}$
and that permutes the basis vectors in such a way that the $(1,1)$ elements of the $2\times 2$ matrices along the diagonal of $\tilde H_F$ are interchanged with the $(2,2)$ elements diagonally above. Here, $\sigma_\pm=(\sigma_x \pm i \sigma_y)/2.$ With respect to this basis, the Floquet Hamiltonian $\hat H_F:=V\tilde H_F V$ reads 
\begin{equation}\label{FinalFloqHam}
\hat H_F=\left( \begin{array}{ccccc}
\ddots&\vdots&\vdots&\vdots&\vdots\\
\cdots&H_{\rm{eff}}+\hbar\omega& \hat H_1&\hat H_2&\cdots\\
\cdots&\hat H_{-1}& H_{\rm{eff}}&\hat H_1&\cdots\\
\cdots&\hat H_{-2}& \hat H_{-1}&H_{\rm{eff}}-\hbar\omega&\cdots\\
\vdots&\vdots&\vdots&\vdots&\ddots\\
\end{array} \right),
\end{equation}
where the $\hat H_i$ are more complicated matrices obtained by interchanging elements of the $\tilde H_j$, which do not need to be defined here. It follows that
\begin{equation}\label{Heff}
H_{\rm{eff}}=\left( \begin{array}{cc}
(\tilde H_0+\hbar\omega)^{2,2} & \tilde H_2^{2,1}  \\
\tilde H_{-2}^{1,2} & (\tilde H_0-\hbar\omega)^{1,1}
\end{array} \right),
\end{equation}
where the superscripts denote a specific element of the corresponding matrix. Using the method from Ref.~\cite{Goldman2014}, which works because $H_{\rm{eff}}$ is smaller than $\hbar\omega$, the corrections to this term can be calculated in terms of commutators of the $\hat H_i$ (for example, the first correction is $[\hat H_{-1},\hat H_1]/\omega$). These terms are all of higher order in $r_0$ and $1/\omega$. For small $\omega$ and/or large shaking amplitudes $r_0$, these higher-order terms become significant. As described in Ref.~\cite{Perez-Piskunow2015}, which the describes the irradiated condensed matter analogue of this system, increasing $r_0$ can lead to phase transitions without additional band inversions, as the off-diagonal blocks become sizeable. For values of $\omega$ around the one at which the two-phonon resonance appears, however, this does not yet occur for the shaking radii $r_0$ that we discuss. This can be seen from a comparison with numerical calculations, which shows that Eq.~\eref{Heff} is sufficient to accurately describe both the gap size and the presence of the topological states; see Fig.~\ref{graph27}(b) for example. 

Although our method can also be used to model the one-phonon resonance, the form of the effective Hamiltonian will be different, as can be seen from the different topological properties connected with this resonance. Higher-phonon resonances occur at low values of $\omega$ and hence require the inclusion of higher-order terms in the effective Hamiltonian. Using the definition of $\tilde H_2$, one obtains
\begin{equation*}
\tilde H_2^{2,1}=\frac{3}{2}J \kappa_2\left(i k_y  -k_x \right)+...,
\end{equation*}
and thus the BHZ Hamiltonian
\begin{equation}\label{BHZHam}
H_{\rm{eff}}=\left( \begin{array}{cc}
M+B k^2a^2 & A(k_x - i k_y)a  \\
A(k_x+i k_y)a & -(M+B k^2a^2)
\end{array} \right),
\end{equation}
where 
\begin{eqnarray}
\frac{M}{J}=\frac{\hbar\omega}{J}-3\kappa_0,\qquad
\frac{B}{J}=\frac{3}{4}\kappa_0 ,\qquad
\frac{A}{J}=-\frac{3}{2}\kappa_2.&\label{A}
\end{eqnarray}
These expressions are correct up to order $(m r_0 \omega a/\hbar)^2$, and agree with the ones derived by Kundu et al. \cite{Kundu2014} upon replacing $e E / \omega$ by $m r_0 \omega$. The presence of the Bessel function of the second kind, $J_2$, through $\kappa_2$ in Eq.~\eref{A}, shows that the opening of a gap at the band inversion is a second-order phonon process. It should be noted that for the NN-hopping $J<0$, one finds that $A$ in Eq.~\eref{A} acquires an additional minus sign, but the spectrum remains unaffected.


\section{Edge states and gap size}

From the effective Hamiltonian in Eq.~\eref{BHZHam}, and following Ref.~\cite{Zhang2010}, one can derive an explicit solution for the edge state in the infinite half-plane. Using perturbation theory to linear order in $k$, the edge states then disperse as 
\begin{equation*}
E_k=\pm Ak=\mp \frac{3}{16}J ka \left(\frac{m r_0 \omega a}{\hbar}\right)^2+...,
\end{equation*}
i.e. the edge states have a velocity quadratic in the frequency $\omega$.

From $H_{\rm{eff}}$, an expression for the gap size $\Delta$ can also be derived, and one obtains
\begin{eqnarray}\label{bandgap}
\Delta =\frac{3}{4}J\sqrt{1-\frac{\hbar\omega}{3J}}\left(\frac{m r_0 \omega a}{\hbar}\right)^2+...
\end{eqnarray}
By substituting the parameter values $\hbar\omega=2.7J$ and $m r_0 \omega^2 a=J$ into Eq.~\eref{bandgap} yields a gap size $\Delta=0.033 J$, which is in good agreement with the numerical results shown in Fig.~\ref{graph27}(b).


\section{Conclusion} In conclusion, we have investigated fermions in a circularly shaken honeycomb optical lattice in the intermediate regime, where the shaking frequency is on the order of the bandwidth $\hbar \omega \approx 3J$. In this particular regime, the system is characterised by a substantial overlap between the Floquet side bands, and a series of band inversions can be created that generally host topological edge states. We have concentrated on the crossing associated with two-phonon resonances, at $\hbar\omega\approx 2.9 J$, and we have shown that the relevant effective continuum model is just the BHZ model for HgTe quantum wells. This allows for an understanding of the transition between the quasi-equilibrium regime and the resonant regime in terms of well-studied effective models, and especially in terms of band inversion, now between adjacent Floquet bands.

Considering that the model Hamiltonian also describes a honeycomb lattice irradiated with circularly polarised light, the question remains whether the discussed effects can be observed in condensed matter. In this case, the phonon resonances become photon resonances, but the prior calculations remain valid, simply by replacing  $m r_0 \omega$ by $eE/ \omega$. A natural candidate would be graphene, but the relevant hopping parameter $J=2.8$~eV and the NN bond length $a=1.4~$\AA~in graphene would require unphysically large frequencies beyond the THz regime, and a very high field strength of $E\approx 5.3\cdot 10^{10}$ V/m. A more promising candidate is a self-assembled honeycomb lattice of CdSe nanocrystals \cite{Kalesaki2014,Bone2014}, which hosts an $s$-band exhibiting a dispersion similar to that of graphene. The hopping parameter in these artificial structures depends on the diameter and the contact area of the nanocrystals. A hopping parameter $J=25$ meV, that is roughly two orders of magnitude smaller than that in graphene, has been theoretically predicted for nanocrystals with a diameter of $3.4$ nm  \cite{Kalesaki2014}. By using light with $E=10^7$ V/m and $\hbar\omega=65$ meV, a gap of $1.5$ meV is obtained for these parameters, which is $6 \%$ of the hopping $J$. In Ref.~\cite{Wang2013}, the Dirac states at the surface of a 3D topological insulator are irradiated by circularly polarised light, and the resulting photon resonance gaps are detected using ARPES. Although thermal excitation of the Dirac electrons is observed, it is possible to measure the Floquet spectrum before the states have been excited away from the bands. This, together with the predicted bandgap, implies that the two-photon resonance should be observable in the recently synthesised artificial superlattices of CdSe nanocrystals \cite{Kalesaki2014,Bone2014}, or in predicted similar structures \cite{Beugeling2015}. 

It would be much more natural to attempt to realise the Hamiltonian in Eq.~\eref{BHZHam} through the use of optical lattices. Honeycomb lattices have been manufactured in the past \cite{Soltan2011}, and they are more promising for two reasons. The first reason for this is the much larger lattice constant, compared to condensed matter systems: since the vector potential enters the Hamiltonian in the combination $eAa,$ this allows for smaller vector potentials. The second reason is that the circular shaking described here creates a vector potential of the form $eA=m r_0 \omega$, as opposed to $E/\omega$, so that increasing the frequency actually increases the vector potential, rather than supressing it. In graphene, for example, the required frequencies suppress the vector potential too strongly, resulting in the necessity of unphysically large electric fields. In contrast, taking a honeycomb optical lattice with NN hopping $J$ and recoil energy $E_r = \hbar^2 k^2 / 2m$, we can rewrite
\begin{equation*}
\frac{aeA}{\hbar}=\frac{a r_0\omega m}{\hbar}=\frac{1}{2}a r_0 \frac{\omega \hbar}{E_r} k^2=2 \pi^2 \frac{r_0}{a}\frac{\hbar \omega}{E_r},
\end{equation*}
To obtain the bandgap derived in the previous section would require shaking by a frequency $\hbar \omega=2.7 J$, at a radius $r_0=aE_r/(140J)$. For potassium atoms loaded in an optical lattice with wavelength $k=1064$~nm, which corresponds to $E_r\approx 4410$~Hz, the shaking would be at several kHz, with a radius of several tens of nm. This suggests that honeycomb optical lattices are a very promising candidate for realizing the topological states discussed here. The richness of shaking protocols, which are a hallmark of optical lattices, together with these encouraging results, promise that the up-and-coming field of atomtronics could be a prime candidate for the experimental investigation of Floquet topological insulators.

The authors would like to thank Daniel Vanmaekelbergh and Michelle Burrello for useful discussions. The work by A.Q. and C.M.S. is part of the D-ITP consortium, a program of the Netherlands Organisation for Scientific Research (NWO) that is funded by the Dutch Ministry of Education, Culture and Science (OCW).\\

\bibliographystyle{jphysicsB}

\end{document}